# EFFECTIVE SOLID ANGLE MODEL AND MONTE CARLO METHOD: IMPROVED ESTIMATIONS TO MEASURE COSMIC MUON INTENSITY AT SEA LEVEL IN ALL ZENITH ANGLES

**Junghyun Bae[1], Stylianos Chatzidakis[1], Robert Bean[1]**

[1]School of Nuclear Engineering, Purdue University, West Lafayette, IN

**ABSTRACT**

*Cosmic muons are highly energetic and penetrative particles and these figures are used for imaging of large and dense objects such as spent nuclear fuels in casks and special nuclear materials in cargo. Cosmic muon intensity depends on the incident angle (zenith angle, $\varphi$), and it is known that $I(\varphi) = I_0 \cos^2 \varphi$ at sea level. Low intensity of cosmic muon requires long measurement time to acquire statistically meaningful counts. Therefore, high-energy particle simulations e.g., GEANT4, are often used to guide measurement studies. However, the measurable cosmic muon count rate changes upon detector geometry and configuration. Here we develop an "effective solid angle" model to estimate experimental results more accurately than the simple cosine-squared model. We show that the cosine-squared model has large error at high zenith angles ($\varphi \geq 60°$), whereas our model provides improved estimations at all zenith angles. We anticipate our model will enhance the ability to estimate actual measurable cosmic muon count rates in muon imaging applications by reducing the gap between simulation and measurement results. This will increase the value of modeling results and improve the quality of experiments and applications in muon detection and imaging.*

## 1. INTRODUCTION

Muons are the most abundant cosmic ray particles on Earth. At sea level, the muon flux is approximately 10,000 per m$^2$ in a minute [1]. Muons are mostly produced by pion decay, $\pi^{\pm} = \mu^{\pm} + \bar{\nu}_{\mu}(or\ \nu_{\mu})$ and they also decay electron/positron and neutrinos with a mean lifetime of 2.2 μsec. A muon is an electron-like lepton but 207 times heavier and unstable [2]. Muons are highly energetic and their large penetration range enable us to utilize them for imaging applications for large and high-density materials. Muon tomography is widely used in geo-tomography [3], cargo scanning [4], and spent nuclear fuel monitoring [5], [6]. Either in a muon scattering tomography or transmission radiography, statistical analysis of muon measurements is used to identify the inside materials via imaging reconstruction algorithm [7], [8]. Hence, it is essential to obtain a statistically meaningful volume of data for high resolution images. Large detectors and long-time measurement enable us to obtain significant muon counts, however an engineering judgement between size, time, and image quality should be considered. Unlike many radiology applications, muon tomography entirely depends on cosmic muons because it is too difficult to artificially produce muons for an imaging purpose. Because cosmic muon intensity is low, Monte Carlo simulation codes, e.g., GEANT4, are performed to study the effect of count rates and measurement time. In practice, a measurable muon counts depend upon the geometry, type, and configuration of detectors. The muon intensity depends on the zenith angle $\varphi$, and it is known to follow a cosine-squared law, such that $I(\varphi) = I_0 \cos^2 \varphi$. In many experimental results, however, this relation is no longer valid at high zenith angles. The cosine-squared model underestimates actual muon counts at high zenith angles, whilst it overestimates counts at low zenith angles [9]. To improve this, we propose a new "effective solid angle" model, which is developed based on the cosine-squared law coupled with the configuration of detector. The effective solid angle model considers the actual measurable solid angle of the detector geometry and setup in the system. In our experiment, two sodium iodine scintillation detectors connected to a coincidence logic gate are installed to count the incident cosmic muon at different zenith angles. Cosmic muons were measured for 24 hours to minimize the day-night count variance [10]. At all zenith angles, our model estimates the results more accurately than that of the simple cosine-squared model. The estimation accuracy is particularly enhanced at high zenith angles ($\varphi \geq 60°$).

## 2. MATERIALS AND METHODS

The sodium Iodide (NaI) scintillation crystal and photomultiplier tube (PMT) are encapsulated within the thin (0.508 mm) aluminum light shielding. The NaI crystal which is located at the bottom of the aluminum housing, has a dimension of 50.8 mm diameter and 50.8 mm height [11], [12]. The photomultiplier base with preamplifier is connected as a module.

### 2.1 Experiment Setup

Cosmic muons at sea level have an average energy of 3-4 GeV. Incoming muon releases about 7 MeV/cm in the NaI detectors by interacting with crystal lattice [13]. The scintillation photon yield depends on the energy loss of muons. Scintillation photons are converted to the photoelectrons hereafter they are multiplied in the PMT. The released energy from a muon to scintillators ultimately results in signals of the preamplifier. Signals are transmitted to the amplifier and they are transformed to the Gaussian-shaped pulse with a thin width. The amplifier gain is set as minimum (0.5×5) in order to suppress the background noise. The deposited energy by a cosmic muon is significant enough to be measured at the minimum amplifier gain. The single channel analyzer (SCA) discriminator level is fixed at the maximum, 10 V, to filter out the background noise. Even two detector systems are operated independently, they share a coincidence logic gate. The coincidence logic gate only accepts signals that occur within 500 ns beyond peak detection from both SCAs of systems [14]. By using the coincidence gate, we were free from the randomly appearing signals. It is less than 1% chance to record random noise simultaneously by two detectors [15]. Consequently, we can selectively detect cosmic muons within a measurable solid angle which is determined by the size of crystals, distance, and alignment.

## 2.2 Signal Output Analysis

Signals from the preamplifier have a prompt rising and delayed tail. The pulse amplitude is proportional to the incoming charges from PMT. Preamplifier pulse depends on the particle energy and bias voltage. Two largest pulses recorded simultaneously are considered as signals by a cosmic muon because typical radiation source does not produce that large pulse in the ambient condition. The pulse height from the amplifier are easily saturated (> 12.2 V) because the preamplifier signals are large enough even when the amplifier gain is fixed as minimum (× 2.5). The amplifier reshapes the pulses to the Gaussian shape and transmits them to the SCA. Squared-pulses are generated from the SCA only if the incoming signal amplitude is greater than the discriminator level (10 V). The output pulse has a width of 500 nsec. It translates that there is a rare chance to measure random background noise by two independent detectors.

## 3. Analytical Model

Based on the projected plane angle and two-detector system, the concept of the "effective solid angle" and model are developed. They are introduced to quantify measurable muon counts in a two-detector system. The important quantities used in the effective solid angle model are described in Fig. 1.

### 3.1 Projected Plane Angle

The projected plane angle is a two-dimensional detectable plane angle denoted as $\theta$ in Fig. 1. Due to the detector geometry and position, the projected plane angle on a detector surface changes along the distance from the centerline of the detector, $r$. For different geometries, an equivalent diameter, $D_e$ of the detector surface can be used:

$$D_e = 2\sqrt{Area/\pi} \qquad (1)$$

Fig. 1 describes an example showing detectable muon traces that pass one point (solid red circle) and its projected plane angle. The projected plane angle depends on the detector surface radius and the distance between two detectors $D$, whereas it is independent to crystal height. The projected plane angle in radian can be expressed as:

$$\theta(r) = \tan^{-1}\left(\frac{r_d + r}{D}\right) + \tan^{-1}\left(\frac{r_d - r}{D}\right) \qquad (2)$$

where $r$ is a distance from the centerline of the detector surface, $r_d$ is a radius of the detector surface. The area averaged projected plane angle $\bar{\theta}$, along the detector surface area $A_d$ is:

$$\bar{\theta} = \frac{\int_{A_d} \theta(r) dA_d}{\int dA_d} \qquad (3)$$

Variance of the maximum, minimum, average, and half-projected plane angle $\gamma$, with a detector surface radius of 2.54 cm and 5.08 cm are estimated using Equations (2), (3) and summarized in Table 1 and plotted in Fig. 2. Projected plane angle becomes smaller when two detectors are placed farther away. If the distance between two detectors is 0, it is identical to a single detector and its projected plane angle becomes 180°. On the contrary, the projected plane angle converges to 0° as the distance between two detectors increases.

### 3.2 Effective Cosmic Muon Solid Angle

The detectable muon intensity is a function of not only the zenith angle, but also of the measurable solid angle. For example, we can observe some muon counts at horizontal placement, $\varphi = 90°$ even though the cosine-squared low predicts zero flux.

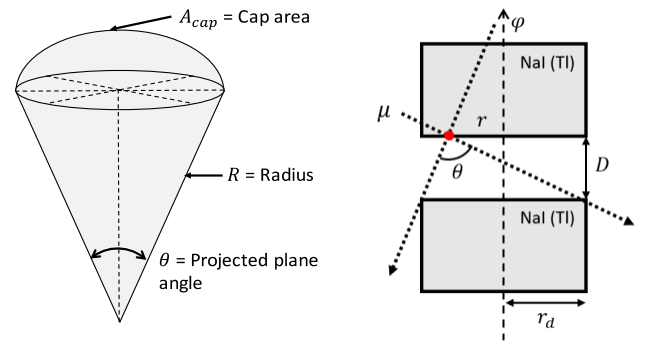

**FIGURE 1:** PROJECTED PLANE ANGLE AND SOLID ANGLE (LEFT). EXAMPLE OF THE DETECABLE MUON TRACES WITH A PROJECTED PLANE ANGLE WITHIN A DETECTION SYSTEM (RIGHT)

**TABLE 1:** MAX, MIN, AVG, AND HALF-PROJECTED PLANE ANGLES $\gamma$, WITH DIFFERENT DETECTOR SURFACE RADII $r_d$, AND DISTANCES $D$

| $D$ [cm] | $r_d$ [cm] | $\theta_{max}$ | $\theta_{min}$ | $\bar{\theta}$ | $\gamma = \bar{\theta}/2$ |
|---|---|---|---|---|---|
| 0 | 2.54 | 180 | 180 | 180 | 90 |
|   | 5.08 | 180 | 180 | 180 | 90 |
| 8 | 2.54 | 35.2 | 32.4 | 33.8 | 16.9 |
|   | 5.08 | 64.8 | 51.8 | 58.0 | 29.0 |
| 9.5 | 2.54 | 29.9 | 28.1 | 29.0 | 14.5 |
|   | 5.08 | 56.3 | 46.9 | 51.4 | 25.7 |
| 11 | 2.54 | 26.0 | 24.8 | 25.4 | 12.7 |
|   | 5.08 | 49.6 | 42.7 | 46.0 | 23.0 |
| ∞ | 2.54 | 0 | 0 | 0 | 0 |
|   | 5.08 | 0 | 0 | 0 | 0 |

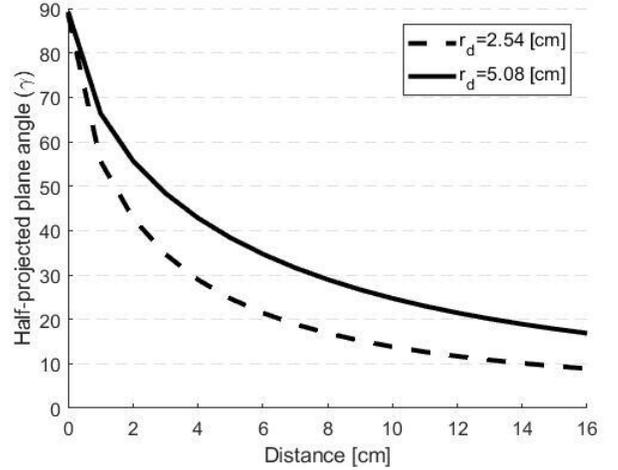

**FIGURE 2:** VARIANCE OF HALF-PROJECTED PLANE ANGLE AS A FUNCTION OF DETECTOR DISTANCE

To consider both detector configurations and zenith angle dependency in the detectable muon count estimation, a cosmic muon effective solid $\Omega_{\text{eff}}$ model is introduced. The measurable intensity of cosmic muons depends on incident zenith angle. It is proportional to the cosine-power of zenith angle,

$$I(\varphi) = I_0(0°) \cos^n \varphi \quad (4)$$

where $\varphi$ is zenith angle, $I_0$ is vertical intensity, and past experiments showed n $\cong$ 2 at sea level [16]. A two-detector installation has a limited measurable solid angle and the muon intensity varies within this solid angle. Assuming there is no muon flux variance in azimuth angle, the solid angle corresponding to the cap area $A_{cap}$ shown in Fig. 1 is given by,

$$\Omega_{cap} = 2\pi \int_0^\gamma \sin\phi\, d\phi \quad (5)$$

where $R$ is a radius, $\gamma$ is the half-projected plane angle. A centerline of a two-detector aligned to zenith angle is changed consequently from 0° to 90°. Each zenith angle with $N$ divisions can be written by:

$$\varphi_i = \frac{\pi}{2N} i \quad (i = 0, 1, 2, \ldots, N) \quad (6)$$

where $\varphi_i$ is the $i^{\text{th}}$ zenith angle. When $R$ represents the muon intensity $I$, the effective intensity area $A_{\text{eff}}$, and the integrated effective intensity solid angle $\Omega'_{\text{eff}}$, coupled with geometry and cosine-power law are derived through Equation (7) to (9).

$$A_{\text{eff}} = 2\pi I_0 \int_{\varphi_i-\gamma}^{\varphi_i+\gamma} I(\phi) \sin\phi\, d\phi \quad (7)$$

$$A_{\text{eff}} = \frac{2\pi I_0^2}{n+1} [\cos^{n+1}(\varphi_i - \gamma) - \cos^{n+1}(\varphi_i + \gamma)] \quad (8)$$

$$\Omega'_{\text{eff}} = \frac{2\pi}{n+1} [\cos^{n+1}(\varphi_i - \gamma) - \cos^{n+1}(\varphi_i + \gamma)] \quad (9)$$

Since $A_{\text{eff}}$ and $\Omega'_{\text{eff}}$ are the integrated effective intensity area and solid angle over the entire azimuth angle, a scaling factor $F$ needs to be introduced to evaluate area of interest:

$$F(i) = \frac{A_{cap}}{A_{2\pi}(i)} = \frac{1 - \cos\gamma}{\cos(\varphi_i - \gamma) - \cos(\varphi_i + \gamma)} \quad (10)$$

where $A_{2\pi}$ is a circular area between $\varphi \pm \gamma$. The general expression of effective solid angle $\Omega_{\text{eff}}$, is:

$$\Omega_{\text{eff}} = \Omega'_{\text{eff}} F(i) \quad (11)$$

when n = 2,

$$\Omega_{\text{eff}}|_{n=2} = \left(\frac{2\pi}{3}\right)[\cos^2(\varphi_i - \gamma) \\ + \cos(\varphi_i - \gamma)\cos(\varphi_i + \gamma) \\ + \cos^2(\varphi_i + \gamma)](1 - \cos\gamma) \quad (12)$$

## 4. MONTE CARLO SIMULATION

In Monte Carlo simulation, arbitrary muon traces are generated on both detector surfaces and connect two points to retrieve muon tracks. They provide the incident muon angles and corresponding relative muon intensities. One hundred muon traces on upper and lower detector surfaces with a distance of 8 cm are simulated. Based on the traces on the detector surfaces, $10^4$ possible muon tracks are reconstructed. Cartesian coordinates of traces on upper and lower surfaces are:

$$(x, y, z)_u = (x_m, y_m, D) \quad m = 1, 2, \ldots, N$$
$$(x, y, z)_l = (x_n, y_n, 0) \quad n = 1, 2, \ldots, N \quad (13)$$

where N is number of muon trace on the detector surface. Reconstructed angles $\theta$, and relative intensity $I$, from point $m$ to point $n$ are,

$$\theta_{m\to n} = \cos^{-1}\left(D/\sqrt{\Delta x^2 + \Delta y^2 + D^2}\right) \quad (14)$$

$$I_{m\to n} = I_0 \cos^2(\theta_{m\to n}) \quad (15)$$

where $\Delta x = x_m - x_n$, $\Delta y = y_m - y_n$. Reconstructed angles and corresponding intensities are expressed as $N \times N$ matrices:

$$\mathbf{\Theta} = \begin{bmatrix} \theta_{1\to 1} & \theta_{1\to 2} & \cdots & \theta_{1\to N} \\ \theta_{2\to 1} & \theta_{2\to 2} & & \theta_{2\to N} \\ \vdots & & & \vdots \\ \theta_{N\to 1} & \theta_{N\to 2} & \cdots & \theta_{N\to N} \end{bmatrix} \quad (16)$$

$$\mathbf{I} = \begin{bmatrix} I_{1\to 1} & I_{1\to 2} & \cdots & I_{1\to N} \\ I_{2\to 1} & I_{2\to 2} & & I_{2\to N} \\ \vdots & & & \vdots \\ I_{N\to 1} & I_{N\to 2} & \cdots & I_{N\to N} \end{bmatrix} \quad (17)$$

At different zenith angles ($\varphi_i$), each muon intensity and mean value are:

$$\mathbf{I}_i = I_0 \cos^2(\mathbf{\Theta} \pm \varphi_i) \quad (18)$$

$$\bar{I} = \frac{1}{N^2} \sum \mathbf{I}_i \quad (19)$$

## 5. RESULTS AND DISCUSSION

Cosmic ray muons are measured for 24 and each measurement is performed at seven different zenith angles, 0°, 15°, ... 75°, 90°. Two detectors are installed in three different distance, 8 cm, 9.5 cm, and 11 cm. The two-detector placement with the distance of 9.5 cm translates to the measurable projected plane angle of 30°. Therefore, it covers all zenith angles by aligning the two-detector centerline with seven zenith angles. The experiment results for 24-hour measurements, effective solid angle model estimation, Monte Carlo estimation, and cosine-squared law are summarized in Table 2, Table 3, and Table 4. Normalized counts and estimations are also included for the comparison with cosine-squared law. At low zenith angles ($\varphi \leq 30°$), cosine-squared law, effective solid angle model, and Monte Carlo method are in good agreement. At $\varphi = 45°$, however, the relative error of the cosine-squared law increases nearly 20% whereas the effective solid angle model and Monte Carlo method have less. The relative error rapidly increases at high zenith angles ($\varphi \geq 60°$) in all models. Cosine-squared model is limited to use for high zenith angles and it is demonstrated in our experiment results. Particularly, the relative error constantly exceeds 50%. Relative error in the effective solid angle model also increases at high zenith angles because detectors begin to detect cosmic muons from the opposite direction. Relative error from both models have a similar trend because the effective solid angle model is mathematically developed based on the cosine-squared model. We demonstrate that the effective solid angle model estimates the cosmic muon intensity with less error, especially at high zenith angles.

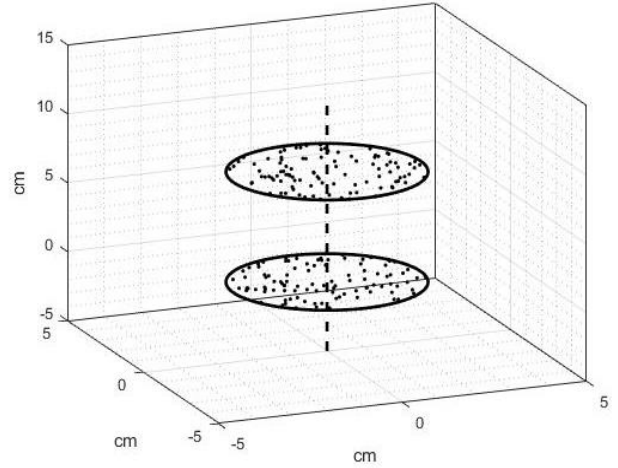

**FIGURE 3:** ARBITRARY GENERATED ONE HUNDRED MUON TRACES (DOTS) ON THE LOWER AND UPPER DETECTOR SURFACES. BASED ON THE TRACES ON THE DETECTOR SURFACE, TEN THOUSAND POSSIBLE MUON TRACKS ARE RECONSTRUCED.

The Monte Carlo simulation has a fine agreement with experimental results within 5% of relative error at most zenith angles. The results of the experiment and normalized muon with different distance estimated by effective solid angle model, cosine-squared model, and Monte Carlo simulation are shown in Fig. 4. The effective solid angle model becomes similar to the cosine-squared model because a detector system becomes a point detector and the solid angle becomes narrower when the distance from two detectors increases.

Measurable cosmic muon intensity variance along detector distances are analyzed using the effective solid angle model. Experiment results and model estimations in four zenith angles, 0°, 30°, 60°, and 90° are shown in Fig. 5. The effective solid angle model estimates with good agreement the measurable muon intensity changes by distances at all angles.

The effective solid angle model is a function of detector surface radius and distance whereas it does not depend on the detector height. Therefore, we can write the effective solid angle with a ratio of detector radius to distance $r_d/D$, and it is shown in Fig. 6. The range $0 < r_d/D < 5$ is corresponding to $0.2r_d < D < 20r_d$, and it represents the practical radius to distance ratio range. The effective solid angle and $r_d/D$ have a linear relation at $r_d/D < 1$. This relation allows us to apply this model to various detector sizes and configurations. $\Omega_{\text{eff}}$ approaches to 2.09 as $r_d/D$ increases ($r_d$ increases or $D$ decreases) because 2.09 radians is the effective solid angle when a single scintillator has a solid angle of $2\pi$ (hemisphere). In addition, $\Omega_{\text{eff}}$ converges to 0 as $r_d/D$ decreases since the measurable solid angle approaches to 0. The experiment results are plotted together with the $\Omega_{\text{eff}}(r_d/D)$ curve in Fig. 6 and they are clearly within the range of $3r_d < D < 5r_d$. In Fig. 6, experiment results also show the linear relation between $r_d/D$ and $\Omega_{\text{eff}}$ which has an agreement with our model estimations.

**TABLE 2**: MEASURED MUON COUNTS (24-HOURS) WITH ERRORS, NORMALIZED MONTE CARLO METHOD, EFFECTIVE SOLID ANGLE MODEL, AND COSINE-SQUARED ESTIMATIONS FOR SEVEN ZENITH ANGLES AT DISTANCE OF 8.0 CM

| Zenith angle ($\varphi$) | Counts for 24 hours | | Monte Carlo Normalization | Effective solid angle | | $\cos^2 \varphi$ |
| --- | --- | --- | --- | --- | --- | --- |
| | Counts | Normalization | | $\Omega_{eff}$ | Normalization | |
| 0 | 1929 | $1.0000 \pm 0.0322$ | 1 | 0.2484 | 1 | 1.000 |
| 15 | 1724 | $0.8937 \pm 0.0296$ | 0.9368 | 0.2324 | 0.9351 | 0.933 |
| 30 | 1472 | $0.7631 \pm 0.0264$ | 0.7638 | 0.1882 | 0.7577 | 0.750 |
| 45 | 1192 | $0.6179 \pm 0.0228$ | 0.5397 | 0.1280 | 0.5154 | 0.500 |
| 60 | 664 | $0.3442 \pm 0.0155$ | 0.3121 | 0.0678 | 0.2731 | 0.250 |
| 75 | 318 | $0.1649 \pm 0.0100$ | 0.1467 | 0.0238 | 0.0957 | 0.067 |
| 90 | 156 | $0.0809 \pm 0.0067$ | 0.0876 | 0.0153 | 0.0615 | 0.000 |

**TABLE 3**: MEASURED MUON COUNTS (24-HOURS) WITH ERRORS, NORMALIZED MONTE CARLO METHOD, EFFECTIVE SOLID ANGLE MODEL, AND COSINE-SQUARED ESTIMATIONS FOR SEVEN ZENITH ANGLES AT DISTANCE OF 9.5 CM

| Zenith angle ($\varphi$) | Counts for 24 hours | | Monte Carlo Normalization | Effective solid angle | | $\cos^2 \varphi$ |
| --- | --- | --- | --- | --- | --- | --- |
| | Counts | Normalization | | $\Omega_{eff}$ | Normalization | |
| 0 | 1533 | $1.0000 \pm 0.0361$ | 1 | 0.2097 | 1 | 1.000 |
| 15 | 1389 | $0.9061 \pm 0.0336$ | 0.9376 | 0.1960 | 0.9347 | 0.933 |
| 30 | 1206 | $0.7867 \pm 0.0303$ | 0.7677 | 0.1586 | 0.7563 | 0.750 |
| 45 | 900 | $0.5871 \pm 0.0247$ | 0.5359 | 0.1075 | 0.5126 | 0.500 |
| 60 | 473 | $0.3085 \pm 0.0162$ | 0.3042 | 0.0564 | 0.2690 | 0.250 |
| 75 | 213 | $0.1389 \pm 0.0102$ | 0.1349 | 0.0190 | 0.0906 | 0.067 |
| 90 | 135 | $0.0881 \pm 0.0079$ | 0.0731 | 0.0106 | 0.0506 | 0.000 |

**TABLE 4**: MEASURED MUON COUNTS (24-HOURS) WITH ERRORS, NORMALIZED MONTE CARLO METHOD, EFFECTIVE SOLID ANGLE MODEL, AND COSINE-SQUARED ESTIMATIONS FOR SEVEN ZENITH ANGLES AT DISTANCE OF 11.0 CM

| Zenith angle ($\varphi$) | Counts for 24 hours | | Monte Carlo Normalization | Effective solid angle | | $\cos^2 \varphi$ |
| --- | --- | --- | --- | --- | --- | --- |
| | Counts | Normalization | | $\Omega_{eff}$ | Normalization | |
| 0 | 1300 | $1.0000 \pm 0.0392$ | 1 | 0.1463 | 1 | 1.000 |
| 15 | 1160 | $0.8923 \pm 0.0360$ | 0.9373 | 0.1367 | 0.9341 | 0.933 |
| 30 | 1008 | $0.7754 \pm 0.0325$ | 0.7650 | 0.1103 | 0.7542 | 0.750 |
| 45 | 742 | $0.5708 \pm 0.0263$ | 0.5291 | 0.0744 | 0.5085 | 0.500 |
| 60 | 365 | $0.2808 \pm 0.0166$ | 0.2929 | 0.0384 | 0.2627 | 0.250 |
| 75 | 209 | $0.1608 \pm 0.0120$ | 0.1197 | 0.0121 | 0.0828 | 0.067 |
| 90 | 114 | $0.0877 \pm 0.0086$ | 0.0558 | 0.0050 | 0.0339 | 0.000 |

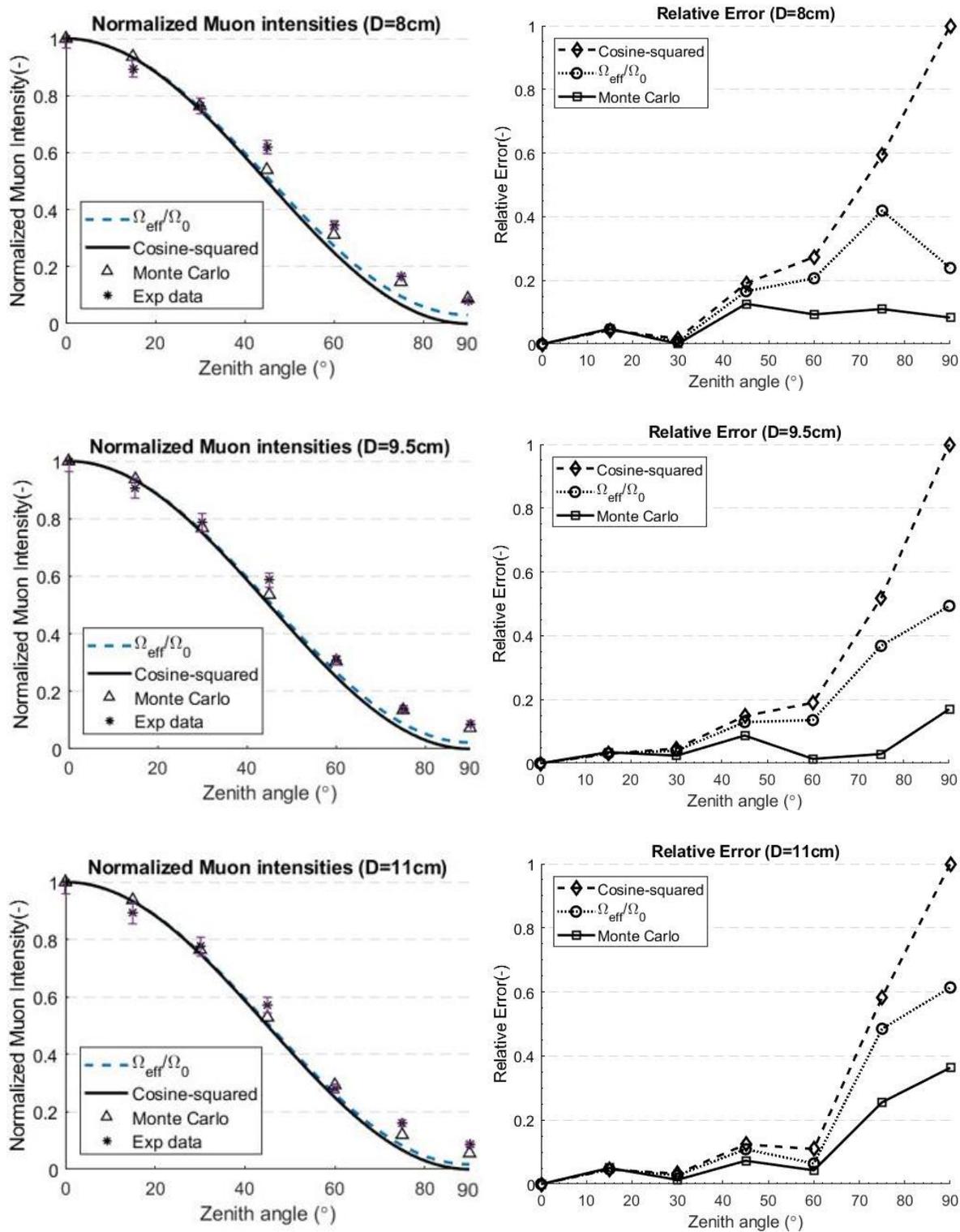

**FIGURE 4:** NORMALIZED MUON INTENSITY ALONG THE ZENITH ANGLES ESTIMATED BY THE EFFECTIVE SOLID ANGLE MODEL (DASHED), COSINE-SQUARED LAW (SOLID), MONTE CARLO METHOD (TRIANGLE), AND EXPERIMENTAL DATA (ASTERISK) AT DISTANCES 8 CM, 9.5 CM, AND 11 CM (LEFT). CORRESPONDING RELATIVE ERRORS WITH THE COSINE-SQUARED LAW (DASHED), EFFECTIVE SOLID ANGLE MODEL (DOTTED), AND MONTE CARLO METHOD (SOLID) ALONG THE ZENITH ANGLES (RIGHT). $\Omega_0$ IS AN EFFECTIVE SOLID ANGLE AT $\varphi = 0°$

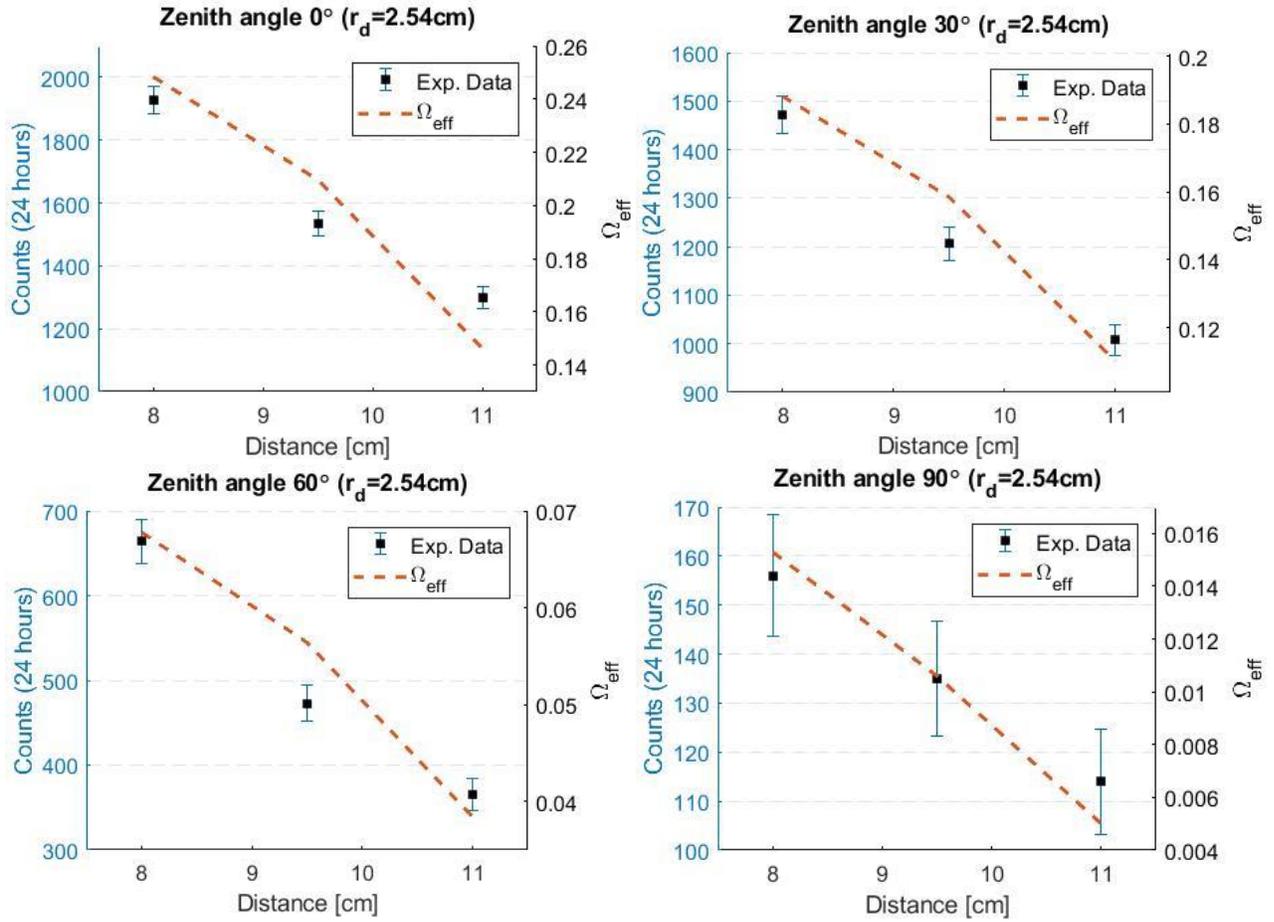

**FIGURE 5:** RELATION BETWEEN MEASURED COSMIC MUON COUNTS (SOLID SQUARE) AND ESTIMATED EFFECTIVE SOLID ANGLES (DASHED) AT 8 CM, 9.5 CM, AND 11 CM WHEN ZENITH ANGLES ARE 0°, 30°, 60°, AND 90°

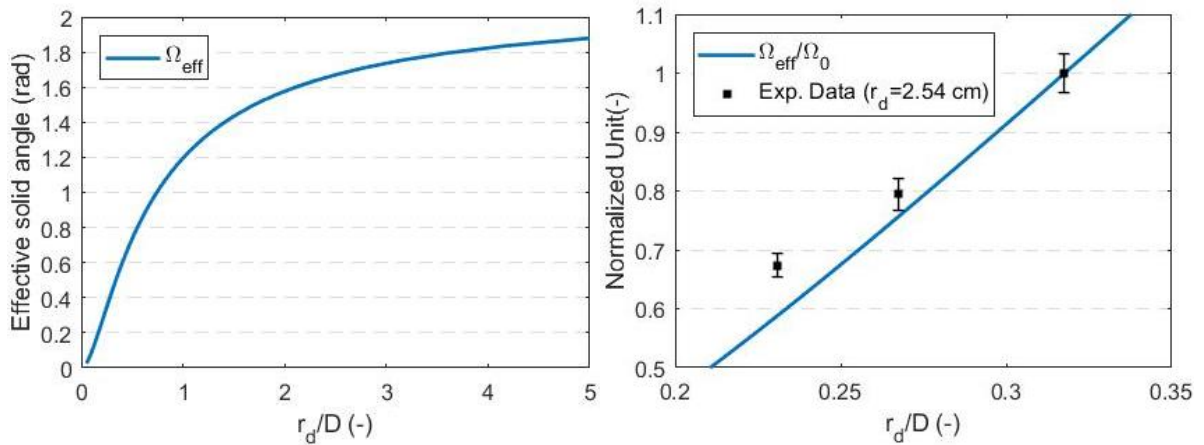

**FIGURE 6:** EFFECTIVE SOLID ANGLE AS A FUNCTION OF A DETECTOR RADIUS TO DISTANCE RATIO $r_d/D$. A GENERAL TREND IN $0.2r_d < D < 20r_d$ (LEFT). EFFECTIVE SOLID ANGLE $\Omega_{\text{eff}}$, APPROACHES 2.09 WHICH REPRESENTS A SINGLE SCINTILLATOR. NORMALIZED EFFECTIVE SOLID ANGLE AND EXPERIMENT DATA WITHIN A RANGE OF $3r_d < D < 5r_d$ (RIGHT)

## 6. CONCLUSION

We discuss the advantages of the proposed effective solid angle model when compared to Monte Carlo simulation and cosine-squared law to estimate incident measurable cosmic muon intensity at sea level. Detectable cosmic muon counts in any system vary depending on detector geometry and setups. Three-dimensional angle that covers detectable muon tracks is termed as a detection solid angle, and the expected muon intensity variances within a detection solid angle is denoted as an effective solid angle. Two sodium iodide scintillation detectors with a coincidence logic gate are installed in our experiment to measure cosmic muon at various zenith angles. The results are normalized to be compared with a cosine-squared estimation, the effective solid angle model, and Monte Carlo simulation. The cosine-squared law shows good agreement at low zenith angles ($\varphi \leq 30°$). However, the error increases significantly for high zenith angles ($\varphi \geq 60°$). On the other hand, the effective solid angle model and Monte Carlo simulation estimate muon intensity with less error compared to the cosine-squared law in all zenith angles and distances.

In addition, the effective solid angle model successfully estimates relative muon intensity at different distances between two detectors. This suggests that the effective solid angle model is a function of a detector radius to distance ratio $r_d/D$, and it can be applied to various sizes and configurations of detection systems. A linear relation is observed when $r_d/D < 1$. This experiment setup has a detector radius to distance ratio range of $0.23 < r_d/D < 0.32$ and results agree with our model estimation showing a linear trend.

Improved measurable cosmic muon intensity estimation capability in all zenith angles enables us to use them horizontally ($\varphi = 90°$) for muon tomographic applications. In addition, the enhanced ability to estimate actual cosmic muon count rates reduces the gap in results between simulation and measurement. We anticipate this will improve modeling quality in muon detection and imaging. For future works, we will elaborate the effective solid angle model to be applied to complex geometries.

## ACKNOWLEDGEMENTS

This research is being performed using funding from the Purdue College of Engineering and the School of Nuclear Engineering.